\documentstyle[12pt,preprint,aps,floats,epsfig,amsbsy,subfigure]{revtex}
\tighten
\newcommand{\bea}{\begin{eqnarray}}
\newcommand{\eea}{\end{eqnarray}}
\newcommand{\beq}{\begin{equation}}
\newcommand{\eeq}{\end{equation}}
\newcommand{\ba}{\begin{array}}
\newcommand{\ea}{\end{array}}
\newcommand{\nn}{\nonumber}
\newcommand{\s}{\hat{s}}
\newcommand{\hm}{\hat{m}_c}

\newcommand{\cb}{\bar{c}}
\renewcommand{\sb}{\bar{s}}

\newcommand{\bra}{\langle}
\newcommand{\ket}{\rangle}

\renewcommand{\a}{\alpha}

\newcommand{\e}{\epsilon}

\renewcommand{\to}{\rightarrow}
\newcommand{\tfrac}[2]{{\frac{#1}{#2}}}
\newcommand{\ts}{}

\newcommand{\eff}{\text{eff}}
\newcommand{\mubeps}{\left( \frac{\mu}{m_b} \right)^{2\epsilon}}

\newcommand{\be}{\begin{equation}}
\newcommand{\ee}{\end{equation}}

\def\Journal#1#2#3#4{{#1} {\bf #2}, #3 (#4)}

\def\NPB{{\em Nucl. Phys.} B}
\def\PLB{{\em Phys. Lett.}  B}
\def\PRL{\em Phys. Rev. Lett.}
\def\PRD{{\em Phys. Rev.} D}

\begin{document}
\thispagestyle{empty}

\preprint{
\noindent
\hfill
\begin{minipage}[t]{6in}
\begin{flushright}
BUTP--01/6     \\
hep-ph/0103087  \\
\vspace*{1.0cm}
\end{flushright}
\end{minipage}
}

\draft

\title{Two-loop virtual corrections to
$\boldsymbol{B \to X_s \ell^+ \ell^-}$  \\ in the standard
model\footnote{Work partially supported by Schweizerischer
Nationalfonds and SCOPES program}}

\vspace{2.0cm}

\author{H.H. Asatryan$^a$, H.M. Asatrian$^a$, C. Greub$^b$ and M. Walker$^b$}

\vspace{2.0cm}

\address{ \vspace{0.3cm}
 a) Yerevan Physics Institute, 2 Alikhanyan Br.,
 375036 Yerevan, Armenia; \\
 b) Institut f\"ur Theoretische Physik, Universit\"at Bern, \\
 CH--3012 Bern, Switzerland.}

\maketitle
\thispagestyle{empty}
\setcounter{page}{0}
%
%=================== ABSTRACT ====================
%
\vspace*{1truecm}
\begin{abstract}
We calculate $O(\alpha_s)$ two-loop virtual 
corrections to the differential
decay width $d\Gamma(B\to X_s \ell^+ \ell^-)/d\s$,
where $\s$ is the invariant
mass squared of the lepton pair, normalized to $m_b^2$.
We also include those contributions from gluon bremsstrahlung
which are needed to cancel infrared and collinear singularities
present in the virtual corrections. Our calculation is 
restricted to the range
$0.05 \le \s \le 0.25$ where the effects from resonances are small.
The new contributions drastically reduce
the renormalization scale dependence of existing results
for $d\Gamma(B\to X_s \ell^+ \ell^-)/d\s$. 
For the corresponding branching ratio
(restricted to the above $\s$-range) 
the renormalization scale uncertainty gets 
reduced from $\sim \pm 13\%$ to $\sim \pm 6.5\%$.
\end{abstract}
%
%==================== INTRODUCTION ==============
%
\vfill
\setlength{\parskip}{1.2ex}
\section{Introduction}
\label{intro}
After the observation of the penguin-induced decay
$B\to X_s \gamma$ \cite{CLEOrare2} and the corresponding exclusive
channels such as $B\to K^*\gamma$ \cite{CLEOrare1},
rare $B$-decays have begun to play an important role in the
phenomenology of particle physics. The measured decay rates are in good
agreement with the standard model (SM) predictions, putting
strong constraints on its various extensions.
Another interesting decay mode in this
context is the inclusive
transition $B \to X_s \ell^+ \ell^-$ ($\ell = e,\mu$).
It has not been observed so far \cite{Glenn:1998gh}, 
but its detection is
expected at the $B$-factories which are currently running.
%While waiting for the experimental results, the theorists should
%make all efforts to reduce the theoretical uncertainties
%in order to get optimal information from future data.
It is known that, unlike for $B\to X_s\gamma $,
large resonant contributions from $\bar{c} c$ intermediate states
come into the game when considering
$B\to X_s \ell^+ \ell^-$.
When
the invariant mass $\sqrt{s}$ of the lepton pair
is close to the mass of a resonance, only model dependent
predictions for these
long distance contributions are available today.
It is therefore unclear
whether integrating the  decay rate over these domains can
reduce the theoretical uncertainty below $\pm 20\%$
\cite{Ligeti:1996yz}.

However, when restricting to regions of $\sqrt{s}$ below the
resonances, the long distance effects are under control.
In particular, all the available studies indicate that for
the region $0.05 \le \hat s = s/m_b^2 \le 0.25$
these non--perturbative effects are below 10\%
\cite{Falk:1994dh}--\cite{Krueger:1996}. Consequently,
the differential decay rate for $B \to X_s \ell^+ \ell^-$
can be predicted precisely in this region
using renormalization group improved
perturbation theory.

It is known that the next-to-leading logarithmic (NLL) result
for the $B \to X_s \ell^+ \ell^-$ decay rate
suffers from a relatively large ($\pm 16\%$) matching 
scale ($\mu_W$)
dependence \cite{Misiak:1993bc,Buras:1995dj}.
To reduce it, next-to-next-to leading (NNLL)
corrections to the Wilson coefficients were 
calculated recently 
by Bobeth et al. \cite{Bobeth:2000mk}.
This required a two-loop matching calculation
of the effective theory to the full
SM theory, followed by a renormalization group treatment of the
Wilson coefficients, using up to three-loop anomalous dimensions
\cite{Bobeth:2000mk,Chetyrkin}.
Including these NNLL corrections to the Wilson coefficients, the
matching scale dependence could be removed to a large extent.

However, this partially NNLL result suffers from a relatively large
($\sim \pm 13\%$)
renormalization scale ($\mu_b$) dependence ($\mu_b \sim O(m_b)$), as
pointed out in ref. \cite{Bobeth:2000mk}. The
aim of the current paper is to reduce this
dependence by  calculating NNLL corrections
to the matrix elements of the
effective Hamiltonian given in the next section.
Our main contribution is the calculation of the
$O(\alpha_s)$ two-loop virtual corrections
to the matrix elements of the operators $O_1$ and $O_2$,
as well  as the $O(\alpha_s)$ one-loop corrections to
$O_7$--$O_{10}$. Also those bremsstrahlung contributions 
are included
which are needed
to cancel infrared and collinear singularities
in the virtual corrections.
The new contributions reduce the renormalization 
scale dependence
from $\sim \pm 13\%$ to $\sim \pm 6.5\%$.

The remainder of this letter is organized as follows.
In Section II we review the theoretical framework.
Our results for the virtual
$O(\alpha_s)$ corrections
to the matrix elements of  the operators $O_1$, $O_2$,
$O_{7}$, $O_8$ and $O_9$ we present in section III.
Section IV is devoted to the bremsstrahlung contributions.
The combined corrections (virtual and bremsstrahlung)
to  $b \to s \ell^+ \ell^-$ are given in section V.
Finally, in section VI we analyze
the invariant mass distribution of the lepton pair in the
range $0.05 \le \s \le 0.25$.
%
%=================== THEORETICAL FRAMEWORK ====================
%
\section{ Theoretical framework}
The most efficient tool for studying weak decays of $B$ mesons is
the effective Hamiltonian technique. For the specific decay channels
$b\to s \ell^+\ell^-$ ($\ell=\mu,e$), the effective Hamiltonian, derived from
the standard model (SM) by integrating
out the $t$-quark and the $W$-boson,
is of the form
\begin{eqnarray}
    \label{Heff}
    {\cal H}_{\eff} =  - \frac{4G_F}{\sqrt{2}} V_{ts}^* V_{tb}
    \sum_{i=1}^{10} C_i \, O_i \quad ,
\end{eqnarray}
where $O_i$ are dimension six operators and
$C_i$ are the corresponding Wilson coefficients.
The operators can be chosen as 
\cite{Bobeth:2000mk}
\begin{equation}
\label{oper}
\begin{array}{rclrcl}
    O_1    & = & (\bar{s}_{L}\gamma_{\mu} T^a c_{L })
                (\bar{c}_{L }\gamma^{\mu} T^a b_{L}) &
    O_2    & = & (\bar{s}_{L}\gamma_{\mu}  c_{L })
                (\bar{c}_{L }\gamma^{\mu} b_{L}) \\ \vspace{0.2cm}
    O_3    & = & (\bar{s}_{L}\gamma_{\mu}  b_{L })
                \sum_q (\bar{q}\gamma^{\mu}  q) &
    O_4    & = & (\bar{s}_{L}\gamma_{\mu} T^a b_{L })
                \sum_q (\bar{q}\gamma^{\mu} T^a q) \\ \vspace{0.2cm}
    O_5    & = & (\bar{s}_L \gamma_{\mu_1} \gamma_{\mu_2}
                \gamma_{\mu_3}b_L)
                \sum_q(\bar{q} \gamma^{\mu_1} \gamma^{\mu_2}\gamma^{\mu_3}q) &
    O_6    & = & (\bar{s}_L \gamma_{\mu_1} \gamma_{\mu_2}
                \gamma_{\mu_3} T^a b_L)
                \sum_q(\bar{q} \gamma^{\mu_1} \gamma^{\mu_2}
                \gamma^{\mu_3} T^a q)    \vspace{0.2cm} \\ \vspace{0.2cm}
    O_7    & = & \frac{e}{g_s^2} m_b (\bar{s}_{L} \sigma^{\mu\nu}
                b_{R}) F_{\mu\nu} &
    O_8    & = & \frac{1}{g_s} m_b (\bar{s}_{L} \sigma^{\mu\nu}
                T^a b_{R}) G_{\mu\nu}^a \\ \vspace{0.2cm}
    O_9    & = & \frac{e^2}{g_s^2}(\bar{s}_L\gamma_{\mu} b_L)
                \sum_\ell(\bar{\ell}\gamma^{\mu}\ell) &
    O_{10} & = & \frac{e^2}{g_s^2}(\bar{s}_L\gamma_{\mu} b_L)
                \sum_\ell(\bar{\ell}\gamma^{\mu} \gamma_{5} \ell) \, ,
\end{array}
\end{equation}
where the subscripts $L$ and $R$ refer to left- and right- handed
components of the fermion fields. We work in the approximation
where the combination
$(V_{us}^* V_{ub})$ of the Cabibbo-Kobayashi-Maskawa (CKM) matrix elements
is neglected; in this case
the CKM structure factorizes, as indicated in eq. (\ref{Heff}).

The factors $1/g_s^2$ in the definition
of the operators $O_7$, $O_9$ and $O_{10}$, as well as the factor
$1/g_s$ present in $O_8$ have been chosen by Misiak \cite{Misiak:1993bc}
in order to simplify
the organization of the calculation: With these definitions,
the one-loop anomalous dimensions (needed for a leading logarithmic
(LL) calculation) of the operators $O_i$ are all proportional to $g_s^2$,
while
two-loop anomalous dimensions (needed for a next-to-leading logarithmic
(NLL) calculation) are proportional to $g_s^4$, etc..

After this important remark we now outline the principal steps
which lead to a LL, NLL, NNLL prediction for the decay amplitude
for $b \to s \ell^+ \ell^-$:
\begin{enumerate}
\item
A matching calculation between the full SM theory and the effective
theory has to be performed
in order to determine the Wilson coefficients $C_i$
at the high scale $\mu_W\sim m_W,m_t$. At this scale, the coefficients
can be worked out in fixed order perturbation theory, i.e. they can be expanded
in $g_s^2$:
\beq
    C_i(\mu_W) = C_i^{(0)}(\mu_W)
    + \frac{g_s^2}{16\pi^2} C_i^{(1)}(\mu_W)
    + \frac{g_s^4}{(16\pi^2)^2} C_i^{(2)}(\mu_W) + O(g_s^6) \, .
\eeq
At LL order, only $C_i^{(0)}$ is needed, at NLL order also $C_i^{(1)}$,
etc.. While the coefficient $C_7^{(2)}$, which is needed for
a NNLL analysis, is known for quite some \cite{matching}, 
$C_{9}^{(2)}$ and  $C_{10}^{(2)}$ have been
calculated only recently \cite{Bobeth:2000mk} (see also \cite{Buchalla:1999ba}).

\item
The renormalization group equation (RGE) has to be solved in order
to get the Wilson
coefficients at the low scale $\mu_b \sim m_b$.
For this RGE step
the anomalous dimension matrix to the relevant
order in $g_s$ is required, as described above.
After these two steps one can decompose the Wilson coefficients
$C_i(\mu_b)$ into a LL, NLL and NNLL part according to
\beq
    \label{wilsondecomplow}
    C_i(\mu_b) = C_i^{(0)}(\mu_b)
    + \frac{g_s^2(\mu_b)}{16\pi^2} C_i^{(1)}(\mu_b)
    + \frac{g_s^4(\mu_b)}{(16\pi^2)^2} C_i^{(2)}(\mu_b) + O(g_s^6) \, .
\eeq
\item
In order to get the decay amplitude,
the matrix elements $\bra s \ell^+ \ell^-|O_i(\mu_b)|b \ket$
have to be calculated. At LL precision, only the operator $O_9$
contributes, as this operator is the only one which at the same time has
a Wilson coefficient starting at lowest order and an explicit
$1/g_s^2$ factor in the definition. Hence, in the NLL precision
QCD corrections (virtual and bremsstrahlung)
to the matrix element of $O_9$ are needed. They have been calculated
a few years ago \cite{Misiak:1993bc,Buras:1995dj}. At NLL precision, also
the other operators start contributing,
viz. $O_7(\mu_b)$ and
$O_{10}(\mu_b)$ contribute at tree-level
and the four-quark operators $O_1,...,O_6$
at one-loop level. Accordingly,
QCD corrections to the latter matrix elements
are needed for a NNLL prediction of the decay amplitude.
\end{enumerate}

As known for a long time \cite{Grinstein:1989}, the formally leading term
$\sim (1/g_s^2) C_9^{(0)}(\mu_b)$ to the amplitude
for $b \to s \ell^+ \ell^-$  is smaller than
the NLL term $\sim (1/g_s^2) [g_s^2/(16 \pi^2)] \, C_9^{(1)}(\mu_b)$.
We adapt our systematics to the numerical situation
and treat the sum of these two terms as a NLL contribution.
This is, admittedly some abuse of language, because the
decay amplitude then starts out with a term which is called NLL.

As pointed out in step 3), $O(\alpha_s)$ QCD corrections to the
matrix elements $\bra s \ell^+ \ell^-|O_i(\mu_b)|b \ket$ have to
be calculated in order to obtain the NNLL prediction for the
decay amplitude. In the present paper we
{\it systematically} evaluate
virtual corrections of order $\alpha_s$
to the matrix elements of $O_1$, $O_2$, $O_7$, $O_8$, $O_9$ and $O_{10}$.
As the Wilson coefficients of the gluonic penguin operators
$O_3,...,O_6$ are much smaller than those of $O_1$ and $O_2$,
we neglect QCD corrections to their matrix elements.
As discussed in more detail later, we also include those
bremsstrahlung diagrams which are needed to cancel
infrared and collinear singularities from the virtual contributions.
The complete bremsstrahlung corrections, i.e. all the finite parts,
however,
will be given elsewhere \cite{AAGW}. We anticipate that the
QCD corrections calculated in the present letter
substantially reduce the scale dependence of the NLL result.
%
%==================== VIRTUAL CORRECTIONS TO THE OPERATORS 
%O1, O2, O7, O8 and O9
% ====================
%
\section{ Virtual corrections to the
operators $O_1$, $O_2$, $O_7$, $O_8$, and $O_9$.}
In this section we present our results for the
virtual $O(\alpha_s)$ corrections induced by the
operators $O_1$, $O_2$, $O_7$, $O_8$, and $O_9$.
Using the naive dimensional regularization (NDR) scheme in
$d=4-2\epsilon$ dimensions,
both, ultraviolet and infrared singularities show up as
$1/\epsilon^n$-poles ($n=1,2$).
The ultraviolet singularities cancel after including the counterterms.
Collinear singularities are regularized by retaining a finite strange
quark mass $m_s$.
They are cancelled together with the infrared singularities
at the level of decay width, taking the bremsstrahlung
process $b \to s \ell^+\ell^- g$ into account.
Gauge invariance implies that the QCD-corrected matrix elements
of the operators $O_i$ can be written as
\beq
    \label{formdef}
        \bra s\ell^+\ell^-|O_i|b\ket =
    \hat{F}_i^{(9)} \bra O_9 \ket_{\text{tree}} +
    \hat{F}_i^{(7)} \bra O_7 \ket_{\text{tree}} \, ,
\eeq
where  $\bra O_9 \ket_{\text{tree}}$ and  $\bra O_7 \ket_{\text{tree}}$
are the tree-level matrix elements of $O_9$ and $O_7$, respectively.
%
%-------------------- VIRTUAL CORRECTIONS TO O1 and O2 --------------------
%
\subsection{Virtual corrections to $O_1$ and $O_2$}
\begin{figure}[t]
    \begin{center}
    \leavevmode
    \includegraphics[height=4cm]{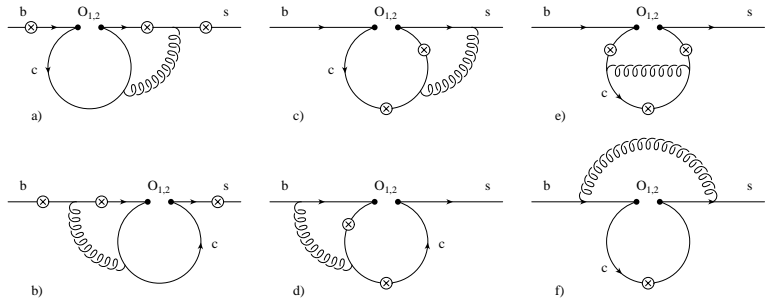}
    \vspace{2ex}
    \caption[f1]{Complete list of two-loop Feynman 
    diagrams for $b \to s \gamma^*$
    associated with the operators $O_1$ and $O_2$.
    The fermions ($b$, $s$ and $c$ quarks) are represented by solid 
    lines;
    the curly lines represent gluons.
    The circle-crosses denote the possible locations for emission 
    of a virtual photon.}
    \label{fig:1}
    \end{center}
\end{figure}

The complete list of Feynman diagrams for the two-loop matrix elements
of the operators $O_1$ and $O_2$ is shown in Fig. 1.
Our calculation follows the line of \cite{Greub:1996tg,Greub:2001sy}
where the contributions of $O_2$ to the processes $B\to X_s\gamma$
and $B\to X_s g$ have been evaluated. There, the results have been found
as expansions in terms of powers and logarithms of the small parameter
$\hat m_c^2=m_c^2/m_b^2$. The central point of the
procedure is to use Mellin-Barnes representations of certain denominators
in the Feynman parameter integrals, as described in detail
in  refs. \cite{Greub:1996tg,Greub:2001sy}.
In the present case, however, we have an additional mass scale: $q^2$,
the invariant mass squared of the lepton pair.
For values of $q^2$ satisfying $\frac{q^2}{m_b^2} < 1$ and 
$\frac{q^2}{4m_c^2} < 1$,
most of the diagrams allow a Taylor series expansion in $q^2$ and can be
calculated in combination with a Mellin-Barnes representation.
This method does not work for the diagram in Fig. \ref{fig:1}a) where
the photon is emitted from the internal line. Instead, we
applied a Mellin-Barnes representation twice.
We will explain this procedure in detail in ref. \cite{AAGW}.
The diagrams in Fig. \ref{fig:1}e) finally, we calculated
using the heavy mass expansion technique \cite{Smirnov}.

Using these methods, the unrenormalized form factors $\hat{F}^{(7,9)}$
of $O_1$ and $O_2$, as defined in eq. (\ref{formdef}), are then obtained in the 
form
\beq
    \hat{F}^{(7,9)} = \sum_{i,j,l,m} c_{ijlm}^{(7,9)}
    \, \hat{s}^i \, \ln^j\!(\hat{s}) \,
    \left(\hat{m}_c^2\right)^l \ln^m\!\left( \hat{m}_c\right),
\eeq
where $\hat{s}=q^2/m_b^2$ and $\hat{m}_c=m_c/m_b$.
$i,j,m$ are non-negative integers and $l=-i,-i+1/2,-i+2/2,....$.
We keep the terms with $i$
and $l$ up to 3, after checking that higher order terms are small
for  $0.05 \le \s \le 0.25$, the range considered in this paper.

The counterterm contributions are of various origin. There are counterterms
due to quark field renormalization,
renormalization of the strong coupling constant $g_s$ and renormalization
of the charm- and bottom- quark masses. We stress that we use
the pole mass definition for both, $m_c$ and $m_b$.
Additionally, we also have to take operator mixing into account.
The corresponding counterterms to the matrix elements
$\bra C_i O_i \ket$ are of the form
\bea
    \label{opmix}
        \bra C_i O_i \ket & = & C_i \sum_j \delta Z_{ij}\bra O_j\ket \, ,\text{ 
with}
    \\
        \delta Z_{ij} & = & \frac{\alpha_s}{4 \pi} \left( a_{ij}^{01} + \frac{1}
{\epsilon}
    a_{ij}^{11}\right) +
        \frac{\alpha_s^2}{(4 \pi)^2}
    \left( a_{ij}^{02} + \frac{1}{\epsilon} a_{ij}^{12} + +
    \frac{1}{\epsilon^2} a_{ij}^{22}\right) +
        O(\alpha_s^3).
\eea
Most of the coefficients $a_{ij}^{lm}$ needed for our calculation are given
in ref. \cite{Bobeth:2000mk}. As some are new, we list those
for $i=1,2$ and $j=1,2,4,7,9,11,12$ that are different from zero:
\beq
    \hat{a}^{11}=
    \left(
    \begin{array}{ccccccc}
        -2      & \frac{4}{3}   & -\frac{1}{9}  & 0 & -\frac{16}{27}    &   
\frac{5}{12}  & \frac{2}{9} \\
        \\
         6      & 0             &  \frac{2}{3}  & 0 & -\frac{4}{9}      &   1   
          & 0
    \end{array}
    \right),\quad
    \begin{array}{lll}
        a^{12}_{17} = -\frac{58}{243}\,, \hspace{0.3cm} &
        a^{12}_{19} = -\frac{64}{729}\,, \hspace{0.3cm} &
        a^{22}_{19} = \frac{1168}{243}\,, \\ \\
        a^{12}_{27} = \frac{116}{81}\,, \hspace{0.3cm} &
        a^{12}_{29} = \frac{776}{243}\,, \hspace{0.3cm} &
        a^{22}_{29} = \frac{148}{81}\,.
    \end{array}
\eeq
$O_{11}$ and $O_{12}$, entering eq. (\ref{opmix}), are evanescent
operators, defined as
\beq
\begin{array}{l}
    O_{11} = \left( \sb_L \gamma_{\mu_1} \gamma_{\mu_2} \gamma_{\mu_3}
             T^a c_L \right)
             \left( \cb_L \gamma^{\mu_1} \gamma^{\mu_2} \gamma^{\mu_3}
             T^a b_L \right) - 16 \, O_1 \\ \\
    O_{12} = \left( \sb_L \gamma_{\mu_1} \gamma_{\mu_2} \gamma_{\mu_3}
             c_L \right)
             \left( \cb_L \gamma^{\mu_1} \gamma^{\mu_2} \gamma^{\mu_3}
             b_L \right) - 16 \, O_2\,.
\end{array}
\eeq
%
%-------------------- BEGIN FIGURES WITH FEYNMAN DIAGRAMS --------
%
\begin{figure}[t]
    \begin{center}
    \leavevmode
    \includegraphics[height=4cm]{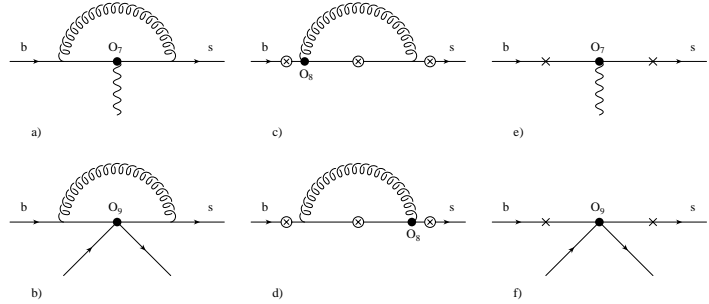}
    \vspace{2ex}
    \caption[f1]{Some Feynman diagrams for
    $b \to s \gamma^*$ or $b \to s \ell^+ \ell^-$
    associated with the operators $O_7$, $O_8$ and $O_9$.
    The circle-crosses denote the possible locations where 
    the virtual photon
    is emitted, while the crosses mark the possible locations
    for gluon bremsstrahlung. See text.}
    \label{fig:2}
    \end{center}
\end{figure}
%
%-------------------- END FEYNMAN DIAGRAMS -------
%
Before we give the result for the renormalized form factors, we remark
that only diagram 1f) (and also its renormalized version)
suffers from infrared and collinear singularities.
As this diagram can easily be combined with diagram
2b) associated with the operator $O_9$, we will take it into
account in the next subsection, when discussing virtual 
corrections to $O_9$.

We decompose the renormalized matrix elements of $O_i$ ($i=1,2$) as 
\beq
\label{o12decomp}
    \bra s \ell^+ \ell^-|C_i^{(0)} O_i | b \ket =
    C_i^{(0)} \left( -\frac{\alpha_s}{4 \pi} \right)
    \left[ F_i^{(9)} \bra \tilde{O}_9 \ket_{\text{tree}} +
        F_i^{(7)} \bra \tilde{O}_7 \ket_{\text{tree}} \right],
\eeq
with $\tilde{O}_9=\frac{\a_s}{4 \pi} \, O_9$ and
     $\tilde{O}_7=\frac{\a_s}{4 \pi} \, O_7$.
The form factors
$F_i^{(9)}$ and $F_i^{(7)}$ read (using $L_\mu = \ln(\mu/m_b)$,
$L_s = \ln (\s)$)
\bea
\ts
    F_1^{(9)} &=&
    \left (-{\tfrac {1424}{729}}+{\tfrac {16}{243}}\,i\pi +
    {\tfrac {64}{27}}\,{ L_c}\right ){ L_\mu}
    -{\tfrac {16}{243}}\,{ L_\mu}\,{ L_s}+
    \left ({\tfrac {16}{1215}}-{\tfrac {32}{135}}\,{{ \hat{m}_c}}^{-2}\right )
    { L_\mu}\,{ \hat{s}}
    \nn \\
    &+&
    \left ({\tfrac {4}{2835}}-{\tfrac {8}{315}}\,{{ \hat{m}_c}}^{-4}\right )
    { L_\mu}\,{{ \hat{s}}}^{2}+
    \left ({\tfrac {16}{76545}}-{\tfrac {32}{8505}}\,
    {{ \hat{m}_c}}^{-6}\right ){ L_\mu}\,{{ \hat{s}}}^{3}
    -{\tfrac {256}{243}}\,{{ L_\mu}}^{2} + f_1^{(9)} \, ,
\eea
\bea
\ts
    F_2^{(9)} &=&
    \left ({\tfrac {256}{243}}-{\tfrac {32}{81}}\,i\pi -
    {\tfrac {128}{9}}\,{ L_c}\right ){ L_\mu}+
    {\tfrac {32}{81}}\,{ L_\mu}\,{ L_s}+
    \left (-{\tfrac {32}{405}}+{\tfrac {64}{45}}\,
    {{ \hat{m}_c}}^{-2}\right ){ L_\mu}\,{ \hat{s}}
    \nn \\ &+&
    \left (-{\tfrac {8}{945}}+{\tfrac {16}{105}}\,
    {{ \hat{m}_c}}^{-4}\right ){ L_\mu}\,{{ \hat{s}}}^{2}+
    \left (-{\tfrac {32}{25515}}+{\tfrac {64}{2835}}\,
    {{ \hat{m}_c}}^{-6}\right ){ L_\mu}\,{{ \hat{s}}}^{3}
    +{\tfrac {512}{81}}\,{{ L_\mu}}^{2}
    +f_2^{(9)} \, ,
\eea
\beq
\ts
    F_1^{(7)} = -{\tfrac {208}{243}}\,{ L_\mu}
    +f_1^{(7)},\quad \quad
    F_2^{(7)} = {\tfrac {416}{81}}\,{ L_\mu}
    +f_2^{(7)} \, .
\eeq
The analytic results for $f_1^{(9)}$, $f_1^{(7)}$, $f_2^{(9)}$, and
$f_2^{(7)}$ (expanded up to $\hat{s}^3$ and $(\hat{m}_c^2)^3$) are 
rather lengthy.
The formulas become relatively short,
however, if we give the charm quark mass dependence in numerical
form (for the characteristic values of 
$\hat{m}_c$=0.27, 0.29 and 0.31).
We write the functions $f_a^{(b)}$ as
\beq
\label{f1decomp}
    f_a^{(b)} = \sum_{i,j} \, k_a^{(b)}(i,j) \,
    \hat{s}^i \, L_s^j  \quad \quad
    \quad (a=1,2;\,b=7,9;\,i=0,...,3;\,j=0,1).
\eeq
The numerical values for the quantities $k_a^{(b)}(i,j)$
are given in Tab. I and II.

\begin{table}[htb]
\label{coeff1}
\begin{center}
\begin{tabular}{| c | rcl | rcl | rcl |}
\hline \ & \multicolumn{3}{c|}{$\hat{m}_c=0.27$} & 
\multicolumn{3}{c|}{$\hat{m}_c=0.29$} 
& \multicolumn{3}{c|}{$\hat{m}_c=0.31$} \\ \hline \hline
$k_1^{(9)}(0,0)$  & $ -12.327$&$+$&$0.13512\,i $  
& $ -11.973$&$+$&$0.16371\,i $
  & $ -11.65$&$+$&$0.18223\,i $
\\
$k_1^{(9)}(0,1)$  & $ -0.080505$&$-$&$0.067181\,i $  & $ 
-0.081271$&$-$&$0.059691\,i $  & $
-0.080959$&$-$&$0.051864\,i $  \\
$k_1^{(9)}(1,0)$  & $ -33.015$&$-$&$0.42492\,i $  & $ 
-28.432$&$-$&$0.25044\,i $
  & $ -24.709$&$-$&$0.13474\,i $
\\
$k_1^{(9)}(1,1)$  & $ -0.041008$&$+$&$0.0078685\,i $  & 
$ -0.040243$&$+$&$0.016442\,i $  & $
-0.036585$&$+$&$0.024753\,i $  \\
$k_1^{(9)}(2,0)$  & $ -76.2$&$-$&$1.5067\,i $  & $ 
-57.114$&$-$&$0.86486\,i $  &
 $ -43.588$&$-$&$0.4738\,i $  \\
$k_1^{(9)}(2,1)$  & $ -0.042685$&$+$&$0.015754\,i $  & $ -0.035191$&$+$&$
0.027909\,i $  & $
-0.021692$&$+$&$0.036925\,i $  \\
$k_1^{(9)}(3,0)$  & $ -197.81$&$-$&$4.6389\,i $  & $ 
-128.8$&$-$&$2.5243\,i $  &
 $ -86.22$&$-$&$1.3542\,i $  \\
$k_1^{(9)}(3,1)$  & $ -0.039021$&$+$&$0.039384\,i $  & $ 
-0.017587$&$+$&$0.050639\,i $  & $
0.013282$&$+$&$0.052023\,i $  \\
\hline \
$k_1^{(7)}(0,0)$  & $ -0.72461$&$-$&$0.093424\,i$ & $ 
-0.68192$&$-$&$0.074998\,i
$ & $ -0.63944$&$-$&$0.05885\,i$
\\
$k_1^{(7)}(0,1)$  &  & 0 &  &  & 0 & & & 0 &  \\
$k_1^{(7)}(1,0)$  & $ -0.26156$&$-$&$0.15008\,i$ & $ -0.23935$&$-$&$0.12289\,i$ 
& $ -0.21829$&$-$&$0.10031\,i$ \\
$k_1^{(7)}(1,1)$  & $ -0.00017705$&$+$&$0.02054\,i$ & $ 
0.0027424$&$+$&$0.019676
\,i$ & $
0.0053227$&$+$&$0.018302\,i$ \\
$k_1^{(7)}(2,0)$  & $ 0.023851$&$-$&$0.20313\,i$ & $ -0.0018555$&$-$&$0.175\,i$ 
& $ -0.022511$&$-$&$0.14836\,i$ \\
$k_1^{(7)}(2,1)$  & $ 0.020327$&$+$&$0.016606\,i$ & $ 0.022864$&$+$&$0.011456\,i
$ & $ 0.023615$&$+$&$0.0059255\,i$
\\
$k_1^{(7)}(3,0)$  & $ 0.42898$&$-$&$0.099202\,i$ & $ 0.28248$&$-$&$0.12783\,i$ &
 $ 0.17118$&$-$&$0.12861\,i$ \\
$k_1^{(7)}(3,1)$  & $ 0.031506$&$+$&$0.00042591\,i$ & $ 0.029027$&$-$&
$0.0082265\, i$ & $ 0.022653$&$-$&$0.0155\,i$
\\
\hline
\end{tabular}
\end{center}
\caption{Coefficients in the decomposition of $f_1^{(9)}$ and $f_1^{(7)}$ for
three values of $\hat{m}_c$ (eq. (\ref{f1decomp})).}
\label{table1}
\end{table}
\begin{table}[htb]
\label{coeff2}
\begin{center}
\begin{tabular}{| c | rcl | rcl | rcl |}
\hline \ & \multicolumn{3}{c|}{$\hat{m}_c=0.27$} & \multicolumn{3}{c|}{$\hat{m}_
c=0.29$} & \multicolumn{3}{c|}{$\hat{m}_c=0.31$} \\ \hline \hline
$k_2^{(9)}(0,0)$  & $ 7.9938$&$-$&$0.81071\,i $  & $ 6.6338$&$-$&$0.98225\,i $  
& $ 5.4082$&$-$&$1.0934\,i $  \\
$k_2^{(9)}(0,1)$  & $ 0.48303$&$+$&$0.40309\,i $  & $ 0.48763$&$+$&$0.35815\,i $
  & $ 0.48576$&$+$&$0.31119\,i $
\\
$k_2^{(9)}(1,0)$  & $ 5.1651$&$+$&$2.5495\,i $  & $ 3.3585$&$+$&$1.5026\,i $  & 
$ 1.9061$&$+$&$0.80843\,i $  \\
$k_2^{(9)}(1,1)$  & $ 0.24605$&$-$&$0.047211\,i $  & $ 0.24146$&$-$&$0.098649\,i
 $  & $ 0.21951$&$-$&$0.14852\,i $
\\
$k_2^{(9)}(2,0)$  & $ -0.45653$&$+$&$9.0402\,i $  & $ -1.1906$&$+$&$5.1892\,i $ 
 & $ -1.8286$&$+$&$2.8428\,i $  \\
$k_2^{(9)}(2,1)$  & $ 0.25611$&$-$&$0.094525\,i $  & $ 0.21115$&$-$&$0.16745\,i 
$  & $ 0.13015$&$-$&$0.22155\,i $
\\
$k_2^{(9)}(3,0)$  & $ -25.981$&$+$&$27.833\,i $  & $ -17.12$&$+$&$15.146\,i $  &
 $ -12.113$&$+$&$8.1251\,i $  \\
$k_2^{(9)}(3,1)$  & $ 0.23413$&$-$&$0.2363\,i $  & $ 0.10552$&$-$&$0.30383\,i $ 
 & $ -0.079692$&$-$&$0.31214\,i $
\\
\hline \
$k_2^{(7)}(0,0)$  & $ 4.3477$&$+$&$0.56054\,i $  & $ 4.0915$&$+$&$0.44999\,i $  
& $ 3.8367$&$+$&$0.3531\,i $  \\
$k_2^{(7)}(0,1)$  & & 0 & & & 0 & & & 0 &  \\
$k_2^{(7)}(1,0)$  & $ 1.5694$&$+$&$0.9005\,i $  & $ 1.4361$&$+$&$0.73732\,i $  &
 $ 1.3098$&$+$&$0.60185\,i $  \\
$k_2^{(7)}(1,1)$  & $ 0.0010623$&$-$&$0.12324\,i $  & 
$ -0.016454$&$-$&$0.11806\, i $  & $
-0.031936$&$-$&$0.10981\,i $  \\
$k_2^{(7)}(2,0)$  & $ -0.14311$&$+$&$1.2188\,i $  & $ 0.011133$&$+$&$1.05\,i $  
& $ 0.13507$&$+$&$0.89014\,i $  \\
$k_2^{(7)}(2,1)$  & $ -0.12196$&$-$&$0.099636\,i $  & 
$ -0.13718$&$-$&$0.068733\, i $  & $
-0.14169$&$-$&$0.035553\,i $  \\
$k_2^{(7)}(3,0)$  & $ -2.5739$&$+$&$0.59521\,i $  & $ -1.6949$&$+$&$0.76698\,i $
  & $ -1.0271$&$+$&$0.77168\,i $
\\
$k_2^{(7)}(3,1)$  & $ -0.18904$&$-$&$0.0025554\,i $  & $ -0.17416$&$+$&$0.049359
\,i $  & $ -0.13592$&$+$&$0.093\,i
$  \\
\hline
\end{tabular}
\end{center}
\caption{Coefficients in the decomposition of $f_2^{(9)}$ and $f_2^{(7)}$ for
three values of $\hat{m}_c$ (eq. (\ref{f1decomp})).}
\label{table2}
\end{table}
%
%-------------------- END TABLES --------------------
%

\subsection{Virtual corrections to the matrix elements of
  $O_7$, $O_8$ and $O_9$}
%
%-------------------- VIRTUAL CORRECTION TO O9 --------------------
%
We first turn to the virtual corrections to the matrix element
of the operator $O_9$, consisting of the vertex correction
shown in Fig. 2b) and of the quark self-energy contributions.
The sum of these corrections
is ultraviolet finite, but suffers from infrared and collinear
singularities. The result can be written as
\beq
    \label{o9decomp}
    \bra s \ell^+ \ell^-|C_9 O_9 | b \ket =
    \tilde{C}_9^{(0)} \left( -\frac{\alpha_s}{4 \pi} \right)
    \left[ F_9^{(9)} \bra \tilde{O}_9 \ket_{\text{tree}} +
        F_9^{(7)} \bra \tilde{O}_7 \ket_{\text{tree}} \right],
\eeq
with $\tilde{O}_9=\frac{\a_s}{4 \pi} \, O_9$ and
$\tilde{C}_9^{(0)}=\frac{4 \pi}{\a_s} \left(C_9^{(0)}+\frac{\a_s}{4 \pi}
C_9^{(1)}\right)$.
The form factors
$F_9^{(9)}$ and $F_9^{(7)}$ read (keeping terms up to order $\s^3$)
\beq
    F_9^{(9)} =  \tfrac{16}{3} + \tfrac{20}{3} \hat{s}  +
        \tfrac{16}{3} \hat{s}^2  + \tfrac{116}{27} \hat{s}^3+f_{\rm{inf}}\, ,
\eeq
\bea
    &&F_9^{(7)} =  -\tfrac{2}{3}\hat{s}\left(1 +
                    \tfrac{1}{2} \hat{s} + \tfrac{1}{3} \hat{s}^2 \right)\, ,
\eea
where the function $f_{\rm{inf}}$ contains the infrared and collinear
singularities. Its explicit form is (using $r=(m_s/m_b)^2$)
\beq
    f_{\rm{inf}}=\frac{\mubeps}{\e} \, \frac{8}{3} \, (1 + \s + \frac{1}{2}\s^2 
+ \frac{1}{3}\s^3)
    + \frac{4}{3} \frac{\mubeps}{\e} \, \ln r + \frac{2}{3} \ln r
    -\frac{2}{3} \ln^2 r \, .
\eeq
At this place, it is convenient to incorporate the renormalized
diagram 1f),
which has not been taken into account so far.
It is easy to see that the two loops factorize into two one-loop
contributions. The charm loop has the Lorentz structure of $O_9$
and can therefore be absorbed into an effective Wilson coefficient:
Diagram 1f) is properly included
by modifying $\tilde{C}^{(0)}_9$ in eq. (\ref{o9decomp}) as follows:
\beq
    \label{c9replacement}
    \tilde{C}^{(0)}_9 \longrightarrow
    \tilde{C}^{(0,\text{mod})}_9=
    \tilde{C}^{(0)}_9 + \left(C_2^{(0)} + 
    \frac{4}{3} C_1^{(0)} \right) 
\, H_0 \, ,
\eeq
where the charm-loop function $H_0$
reads (in expanded form)
\beq
    H_0 = \frac{1}{2835} \left[
    -1260 + 2520 \ln (\mu/m_c) + 252 \s \hm^{-2} +  
    27 \s^2 \hm^{-4} +
    4 \s^3 \hm^{-6} \right] \, .
\eeq
In the context of virtual corrections
also the $O(\e)$-part of this loop function is needed.
We neglect it here
since it will drop out in combination with gluon bremsstrahlung.
Note that $H_0=h(\hat{m}_c^2,\s)+8/9 \, \ln(\mu/m_b)$,
with $h$ defined in \cite{Buras:1995dj,Bobeth:2000mk}.
%
%---------- Corrections to the matrix elements of O7
%

We now turn to the virtual corrections to the matrix element
of the operator $O_7$, consisting of the vertex-
(Fig. 2a) and self-energy corrections.
The sum of these diagrams is ultraviolet singular. After
renormalization, the result can be written as
\beq
    \label{o7decomp}
    \bra s \ell^+ \ell^-|C_7 O_7 | b \ket =
    \tilde{C}_7^{(0)} \left( -\frac{\alpha_s}{4 \pi} \right)
    \left[ F_7^{(9)} \bra \tilde{O}_9 \ket_{\text{tree}} +
        F_7^{(7)} \bra \tilde{O}_7 \ket_{\text{tree}} \right],
\eeq
with $\tilde{O}_7=\frac{\a_s}{4 \pi} \, O_7$ and
$\tilde{C}_7^{(0)}=C_7^{(1)}$.
The form factors $F_7^{(9)}$ and $F_7^{(7)}$ read
\beq
\ts
    F_7^{(9)} = -\tfrac{16}{3}\left(1 + \tfrac{1}{2} \hat{s}
                +\tfrac{1}{3} \hat{s}^2 + \tfrac{1}{4} \hat{s}^3 \right)\, ,
\eeq
\beq
\ts
    F_7^{(7)} =
    {\tfrac{32}{3}}\,{ L_\mu}
    +\tfrac{32}{3} + 8 \hat{s} + 6 \hat{s}^2 + \tfrac{128}{27} \hat{s}^3+f_{\rm{
inf}}\, .
\eeq
Note that for these expressions the
{\it pole mass for $m_b$} has to be used at lowest order.

%
%---------- Corrections to the matrix elements of O8 ----------
%
Finally, we give the result for the renormalized corrections to
the matrix elements of $O_8$. The corresponding diagrams are shown
in Fig. 2c) and 2d). One obtains:
\beq
    \label{o8decomp}
    \bra s \ell^+ \ell^-|C_8 O_8 | b \ket =
    \tilde{C}_8^{(0)} \left( -\frac{\alpha_s}{4 \pi} \right)
    \left[ F_8^{(9)} \bra \tilde{O}_9 \ket_{\text{tree}} +
        F_8^{(7)} \bra \tilde{O}_7 \ket_{\text{tree}} \right],
\eeq
with $\tilde{C}_8^{(0)}=C_8^{(1)}$.
The form factors $F_8^{(9)}$ and $F_8^{(7)}$ read (in expanded form)
\bea
    F_8^{(9)} & = &
    \tfrac{104}{9} - \tfrac{32}{27} \pi^2 +
    \left(\tfrac{1184}{27} - \tfrac{40}{9} \pi^2\right) \hat{s} +
    \left(\tfrac{14212}{135} - \tfrac{32}{3} \pi^2 \right)\hat{s}^2
    \nn \\ & + &
    \left(\tfrac{193444}{945} - \tfrac{560}{27} \pi^2 \right) \hat{s}^3 +
    \tfrac{16}{9} L_s \left( 1 + \hat{s} + \hat{s}^2 +  
    \hat{s}^3 \right)\, ,
\eea
\bea
    F_8^{(7)} & = &
    - \tfrac {32}{9}\, L_\mu +
    \tfrac{8}{27} \pi^2 - \tfrac{44}{9} - \tfrac{8}{9} i \pi +
    \left(\tfrac{4}{3} \pi^2 - \tfrac{40}{3} \right) \hat{s}
    + \left(\tfrac{32}{9} \pi^2 - \tfrac{316}{9} \right) \hat{s}^2
    \nn \\ & + &
    \left(\tfrac{200}{27} \pi^2 - \tfrac{658}{9} \right) \hat{s}^3 -
    \tfrac{8}{9}\, L_s  \left( \hat{s} + \hat{s}^2 + \hat{s}^3 \right)\, .
\eea
%
%==================== BREMSSTRAHLUNG CORRECTIONS ===============
%
\section{Bremsstrahlung corrections}
We stress  that in the present paper only those
bremsstrahlung diagrams are taken into account which are needed to cancel
the infrared and collinear singularities from the virtual
corrections. All other bremsstrahlung contributions
(which are finite), will be given elsewhere \cite{AAGW}.

It is known \cite{Misiak:1993bc,Buras:1995dj}
that the contribution to the inclusive decay width coming from
the interference between the tree-level and the one-loop matrix elements of
 $O_9$ (Fig. 2b)) and  from the corresponding bremsstrahlung
corrections (Fig. 2f)),
can be written in the form
\begin{eqnarray}
    \label{gamma99}
    \frac{d\Gamma_{99}}{d\hat s}=\left(\frac{\alpha_{em}}{4\pi}\right)^2
    \frac{G_F^2 m_{b,pole}^5\left|V_{ts}^*V_{tb}\right|^2}
    {48\pi^3}(1-\hat s)^2
    \left (1+2\hat s\right)\left
    (2\left (\widetilde C_9^{(0)}\right)^2\frac{\alpha_s}
    {\pi}\omega_9(\hat{s})\right ),
\end{eqnarray}
where $\widetilde C_9^{(0)}=\frac {4 \pi}{\alpha_s}
\left (C_9^{(0)}+\frac {\alpha_s}{4 \pi} C_9^{(1)}
\right )$; the function $\omega_9(\hat{s})\equiv\omega(\hat{s})$,
which contains information on virtual and bremsstrahlung corrections,
can be found in \cite{Misiak:1993bc,Buras:1995dj}.
%\begin{eqnarray}
%\nonumber
%&&\omega_9(\hat{s})=-\frac{4}{3}Li_2(\hat{s})-
%\frac{2}{3}\ln(1-\hat{s})\ln(\hat{s})-\frac{2}{9}\pi^2
%-\frac{5+4\hat{s}}{3(1+2\hat{s})}\ln(1-\hat{s})\\
%&&-\frac{2\hat{s}(1+\hat{s})(1-2\hat{s})}{3(1-\hat{s})^2
%(1+2\hat{s})}\ln(\hat{s})+\frac{5+9\hat{s}-6\hat{s}^2}
%{6(1-\hat{s})(1+2\hat{s})}
%\end{eqnarray}
%
Replacing $\tilde{C}_9^{(0)}$ by $\tilde{C}_9^{(0,\text{mod})}$
(see eq. (\ref{c9replacement})) in
eq. (\ref{gamma99}), diagram 1f) and
the corresponding bremsstrahlung corrections are automatically included.

Similarly, the contribution to the decay width from the
interference between the tree-level and the
one-loop matrix element of $O_7$ (Fig. 2a), combined
with the corresponding bremsstrahlung corrections
shown in Fig. 2e), can be written as
\begin{eqnarray}
    \frac{d\Gamma_{77}}{d\hat s}=\left(\frac{\alpha_{em}}{4\pi}\right)^2
    \frac{G_F^2 m_{b,pole}^5\left|V_{ts}^*V_{tb}\right|^2}
    {48\pi^3}(1-\hat s)^2
    4\left(1+2/\hat s\right)
    \left(2\left (\widetilde C_7^{(0)}\right)^2\frac{\alpha_s}
    {\pi}\omega_7(\s)\right ),
\end{eqnarray}
where $\widetilde C_7^{(0)}=C_7^{(1)}$.
The function $\omega_7(\s)$, which is new, reads
\begin{eqnarray}
    \omega_7(\hat s) &=&
    - \frac{8}{3}\,\ln \left( \frac{\mu}{m_b} \right) -
    \frac{4}{3}\, \mbox{Li}(\s)
    - \frac{2}{9} \,{\pi }^{2}
    - \frac{2}{3}\, \ln(\s) \ln(1-\s) \\
    \nonumber
    &&
    - \frac{1}{3}\, \frac{8 + \s}{2 + \s} \ln (1-\s) -
    \frac{2}{3}\,
    \frac {\s \left( 2 - 2\,\s - \s^2 \right)}
    { \left( 1 - \s \right)^{2} \left( 2 + \s \right)}
    \ln(\s) -
    \frac{1}{18}\,
    \frac {16 - 11\,\s - 17\,\s^{2}}
    {\left( 2 + \s \right) \left( 1 - \s \right)} \, .
\end{eqnarray}

Finally, one observes that also the
interference between the tree-level matrix
element of $O_7$ and the one-loop
matrix element of $O_9$ (and vice versa)
lead to an infrared singular contribution to the decay width.
We combined it with the corresponding
bremsstrahlung terms coming
from the interference of diagrams 2e) and  2f). The result reads
\begin{eqnarray}
    \frac{d\Gamma_{79}}{d\hat s}=
    \left(\frac{\alpha_{em}}{4\pi}\right)^2
    \frac{G_F^2 m_{b,pole}^5\left|V_{ts}^*V_{tb}\right|^2}
    {48\pi^3}(1-\hat s)^2
    12\cdot 2 \frac{\alpha_s}
    {\pi}\omega_{79}(\s)
    \mbox{Re}\left (\widetilde C_7^{(0)}
    \widetilde C_9^{(0)}\right ) \, .
\end{eqnarray}
For the function
$\omega_{79}(\hat{s})$, which also is new, we obtain
\begin{eqnarray}
    \omega_{79}(\s) &=&
    - \frac{4}{3}\,\ln \left (\frac {\mu}{m_b}\right)
    -\frac{4}{3}\, \mbox{Li}(\s) -
    \frac{2}{9}\,{\pi }^{2} -
    \frac{2}{3}\,\ln(\s) \ln(1 - \s) \\
    \nonumber
    &&
    - \frac{1}{9}\,\frac{2 + 7\s}{\s} \ln (1-\s)
    -\frac{2}{9}\, \frac{\s \left(3 - 2\,\s \right)}
    {\left( 1 - \s \right)^{2}} \ln(\s)
    + \frac{1}{18}\, \frac{5- 9\,\s}{1-\s} \, .
\end{eqnarray}
%
%==================== CORRECTION TO THE DECAY WIDTH FOR B->Xs l+ l-
%
\section{Corrections to the decay width for $b \to X_s \ell^+ \ell^-$}
In this section we combine  the virtual corrections
calculated in section III and the bremsstrahlung contributions
discussed in section IV and study their influence on the decay width
$d\Gamma(b \to X_s \ell^+ \ell^-)/d\s$. In the literature
(see e.g. \cite{Bobeth:2000mk}), this decay width is usually written as
\begin{eqnarray}
\label{rarewidth}
    &&\frac{d\Gamma(b\to X_s \ell^+\ell^-)}{d\hat s}=
    \left(\frac{\alpha_{em}}{4\pi}\right)^2
    \frac{G_F^2 m_{b,pole}^5\left|V_{ts}^*V_{tb}\right|^2}
    {48\pi^3}(1-\hat s)^2 \times \nn \\
    &&\left ( \left (1+2\hat s\right)
    \left (\left |\widetilde C_9^{\eff}\right |^2+
    \left |\widetilde C_{10}^{\eff}\right |^2 \right )
    + 4\left(1+2/\hat s\right)\left
    |\widetilde C_7^{\eff}\right |^2+
    12 \mbox{Re}\left (\widetilde C_7^{\eff}
    \widetilde C_9^{\eff*}\right ) \right ) \, ,
\end{eqnarray}
where the contributions calculated so far have been
absorbed into the effective Wilson coefficients
$\tilde{C}_7^{\eff}$, $\tilde{C}_9^{\eff}$ and 
$\tilde{C}_{10}^{\eff}$.
It turns out that also the new contributions calculated in
the present paper can be absorbed into these coefficients. 
Following
as closely as possible the 'parametrization' given recently 
by Bobeth et al.
\cite{Bobeth:2000mk}, we write
\begin{eqnarray}
    \label{effcoeff}
    \nonumber
    \widetilde C_9^{\eff}&=&\left (1+\frac{\alpha_s(\mu)}{\pi}
    \omega_9 (\hat{s})\right )
    \left (A_9+T_9 \, h (\hat m_c^2,
    \hat{s})+U_9 \, h (1,\hat{s})+
    W_9 \, h (0,\hat{s})\right)\\
    && -\frac{\alpha_{s}(\mu)}{4\pi}\left(C_1^{(0)} F_1^{(9)}+
    C_2^{(0)} F_2^{(9)}+
    A_8^{(0)} F_8^{(9)}\right)\\
    \nonumber
    \widetilde C_7^{\eff}&=&\left (1+\frac{\alpha_s(\mu)}{\pi}
    \omega_7 (\hat{s})\right )A_7
    -\frac{\alpha_{s}(\mu)}{4\pi}\left(C_1^{(0)} F_1^{(7)}+
    C_2^{(0)} F_2^{(7)}+A_8^{(0)} F_8^{(7)}\right)\\
    \nonumber
    \widetilde C_{10}^{\eff}&=& \left (1+
    \frac{\alpha_s(\mu)}{\pi}
    \omega_9 (\hat{s})\right ) A_{10} \, ,
\end{eqnarray}
where the expressions for $h (\hat m_c^2,\hat{s})$
and $\omega_9(\s)$ are given in \cite{Bobeth:2000mk}.
The quantities $\omega_7(\s)$ and $F_{1,2,8}^{(7,9)}$, on the other
hand, have been calculated in the present paper.
We take the numerical values for
$A_7$, $A_9$, $A_{10}$, $T_9$, $U_9$, and $W_9$
from \cite{Bobeth:2000mk}, while $C_1^{(0)}$, $C_2^{(0)}$ and
$A_8^{(0)}=\tilde{C}_8^{(0,\eff)}$ are taken from
\cite{Greub:2001sy}.
For completeness
we list them  in Tab. III.
\begin{table}[htb]
\label{coeff3}
\begin{center}
\begin{tabular}{| l | c | c | c |}
&$\mu=2.5$ GeV& $\mu=5$ GeV  & $\mu=10$ GeV\\ \hline
$\alpha_s$ & $0.267$&  $0.215$ & $  0.180$\\ \hline
$C_1^{(0)}$ & $-0.697$ & $-0.487$ & $-0.326$\\ \hline
$C_2^{(0)}$ & $1.046 $& $ 1.024$ &  $ 1.011$\\ \hline
$(A_7^{(0)},A_7^{(1)})$ & $ (-0.360, 0.031)$ & $(-0.321,0.019)$   & $ (-0.287,0.
008)$\\ \hline
$A_8^{(0)}$ &$ -0.164 $& $ -0.148$& $ -0.134$\\ \hline
$(A_9^{(0)},A_9^{(1)})$ & $ (4.241,-0.170) $& $(4.129, 0.013)$ & $ (4.131, 0.155
)$ \\ \hline
$(T_9^{((0))},T_9^{(1)})$ & $ (0.115, 0.278) $& $ (0.374,0.251)$ & $ (0.576,0.23
1)$ \\ \hline
$(U_9^{(0)},U_9^{(1)})$ & $ (0.045,0.023)$ & $ (0.032,0.016)$ & $ (0.022,0.011)$
 \\ \hline
$(W_{9}^{(0)},W_{9}^{(1)})$ & $(0.044,0.016)$ & $ (0.032,0.012)$ & $ (0.022,0.00
9)$ \\ \hline
$(A_{10}^{(0)},A_{10}^{(1)})$ & $ (-4.372,0.135)$ &  $(-4.372,0.135)$ &  $ (-4.3
72,0.135)$ \\
\hline
\end{tabular}
\end{center}
\caption{Coefficients appearing in eq. (\ref{effcoeff}) for $\mu = 2.5$ GeV,
$\mu =5$ GeV and $\mu = 10$ GeV. For $\alpha_s(\mu)$
(in the $\overline{\mbox{MS}}$ scheme) we used the
two-loop expression with 5 flavors and $\a_s(m_Z)=0.119$.
The entries correspond to the pole top quark mass
$m_t= 174 $ GeV. The superscript $(0)$ refers to lowest order
quantities and while the superscript $(1)$ denotes the correction terms
of order $\alpha_s$.}
\label{table3}
\end{table}

When calculating the decay width (\ref{rarewidth}), we retain only
terms linear in $\alpha_s$ (and thus in  $\omega_9$ and $\omega_7$)
in $|\widetilde C_9^{\eff}|^2$ and
$|\widetilde C_7^{\eff}|^2$.
In the interference term
$\mbox{Re} \left (\widetilde C_7^{\eff} \widetilde C_9^{\eff*}
\right )$ too, we keep only terms linear in $\alpha_s$. By construction,
one has to make the replacements
$\omega_9 \to \omega_{79}$ and $\omega_7\to \omega_{79}$ in this term.

Our results include all the relevant virtual corrections
and singular bremsstrahlung contributions. There exist
additional bremsstrahlung terms coming e.g. from one-loop
$O_1$ and $O_2$ diagrams in which both, the virtual photon and the gluon
are emitted from the charm quark line.
These contributions do not induce additional renormalization scale 
dependence
as they are ultraviolet finite. Using our experience from $b \to s \gamma$
and $b \to s g$, these contributions are not expected to be large.

\section{Numerical results}
The decay width in eq. (\ref{rarewidth}) has a large 
uncertainty due to the factor
$m_{b,pole}^5$. Following common practice, we consider 
the ratio
\begin{equation}
R_{\text{quark}}(\s)=
\frac{1}{\Gamma(b \to X_c e\bar{\nu})}\frac{d\Gamma(b\to
X_s \ell^+ \ell^-)}{d\hat s} \, ,
\end{equation}
in which the factor $m_{b,pole}^5$ drops out. The explicit
expression for the semi-leptonic decay width $\Gamma(b \to X_c e \nu_e)$
can be found e.g. in \cite{Bobeth:2000mk}.

We now turn to the numerical results for $R_{\text{quark}}(\s)$ for
$0.05 \le \s \le 0.25$. In Fig. 3a we investigate  the dependence of
$R_{\text{quark}}(\s)$ on the renormalization scale $\mu$. The solid
lines are obtained by including the new NNLL contributions
as explained in detail in section V. The three solid lines correspond
to $\mu=2.5$ GeV (lower line), $\mu=5$ GeV (middle line) 
and $\mu=10$ GeV (upper line).
The three dashed lines (again $\mu=2.5$ GeV for the lower, 
$\mu=5$ GeV for the middle
and $\mu=10$ GeV for the upper curve), on the other hand,
show the results without the new NNLL corrections, i.e., 
they include the NLL
results combined with the NNLL corrections to the matching conditions
as obtained by Bobeth et al. \cite{Bobeth:2000mk}.
{}From this figure we conclude that the 
renormalization scale dependence
gets reduced by more than a factor of 2. Only for small values of $\s$
($\s \sim 0.05$), where the NLL $\mu$-dependence is small already, the
reduction factor is smaller. For the integrated quantity we obtain
\beq
\label{Rint}
R_{\text{quark}} = \int_{0.05}^{0.25} \, d\s \, R_{\text{quark}}(\s)
= (1.25 \pm 0.08 ) \times 10^{-5}
\, ,
\eeq
where the error is obtained by varying $\mu$ between 2.5 GeV and 10 GeV.
Before our corrections, the result was
$R_{\text{quark}}=(1.36 \pm 0.18)\times 10^{-5}$ 
\cite{Bobeth:2000mk}. In other
words, the renormalization scale dependence got 
reduced from $\sim \pm 13\%$ to
$\sim \pm  6.5\%$.

Among the errors on  $R_{\text{quark}}(\s)$ which are due to
the uncertainties in the input parameters, the one induced by
$\hat{m}_c=m_c/m_b$ is known to be the largest. We therefore
show in Fig. 3b the dependence of $R_{\text{quark}}(\s)$ on
$\hat{m}_c$. Comparing
Fig. 3a with Fig. 3b, we find that the uncertainty due to $\hat{m}_c$ is
somewhat larger than the left-over 
$\mu$-dependence at the NNLL level. For the
integrated quantity $R_{\text{quark}}$
we find an uncertainty of $\pm 7.6\%$ due to $\hat{m}_c$.

To conclude: We have calculated virtual 
corrections of order $\a_s$ to the
matrix elements of $O_1$, $O_2$, $O_7$, $O_8$, $O_9$ and $O_{10}$.
We also took into account those bremsstrahlung corrections which cancel
the infrared and collinear singularities in the virtual corrections.
The renormalization scale dependence of $ R_{\text{quark}}(\s) $
gets reduced by more than a factor of 2. 
The calculation of the remaining
bremsstrahlung contributions 
(which are expected to be rather small) and
a more detailed numerical analysis are in 
progress \cite{AAGW}.

%
%--------------- BEGIN PLOTS ---------------
%
\begin{figure}
%    \subfigure[$\mu$-dependence of $R_{\text{quark}}(\s)$]
    {
        \hspace{0.5cm}
        \includegraphics[width=7.5cm, bb=10 205 592 540]{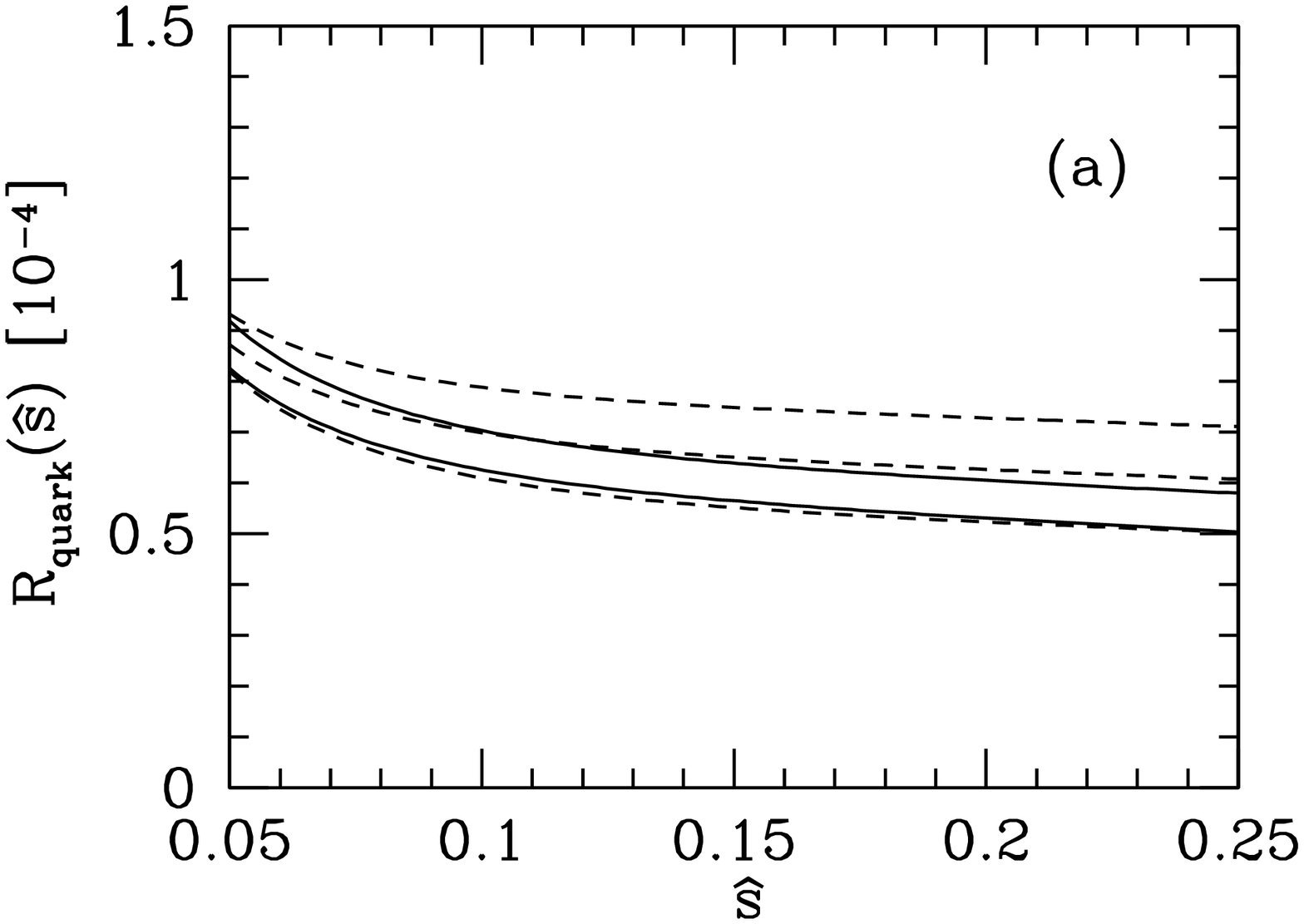}
    }
%    \subfigure[$\hat{m_c}$-dependence of $R_{\text{quark}}(\s)$]
    {
        \includegraphics[width=7.5cm, bb=10 205 592 540]{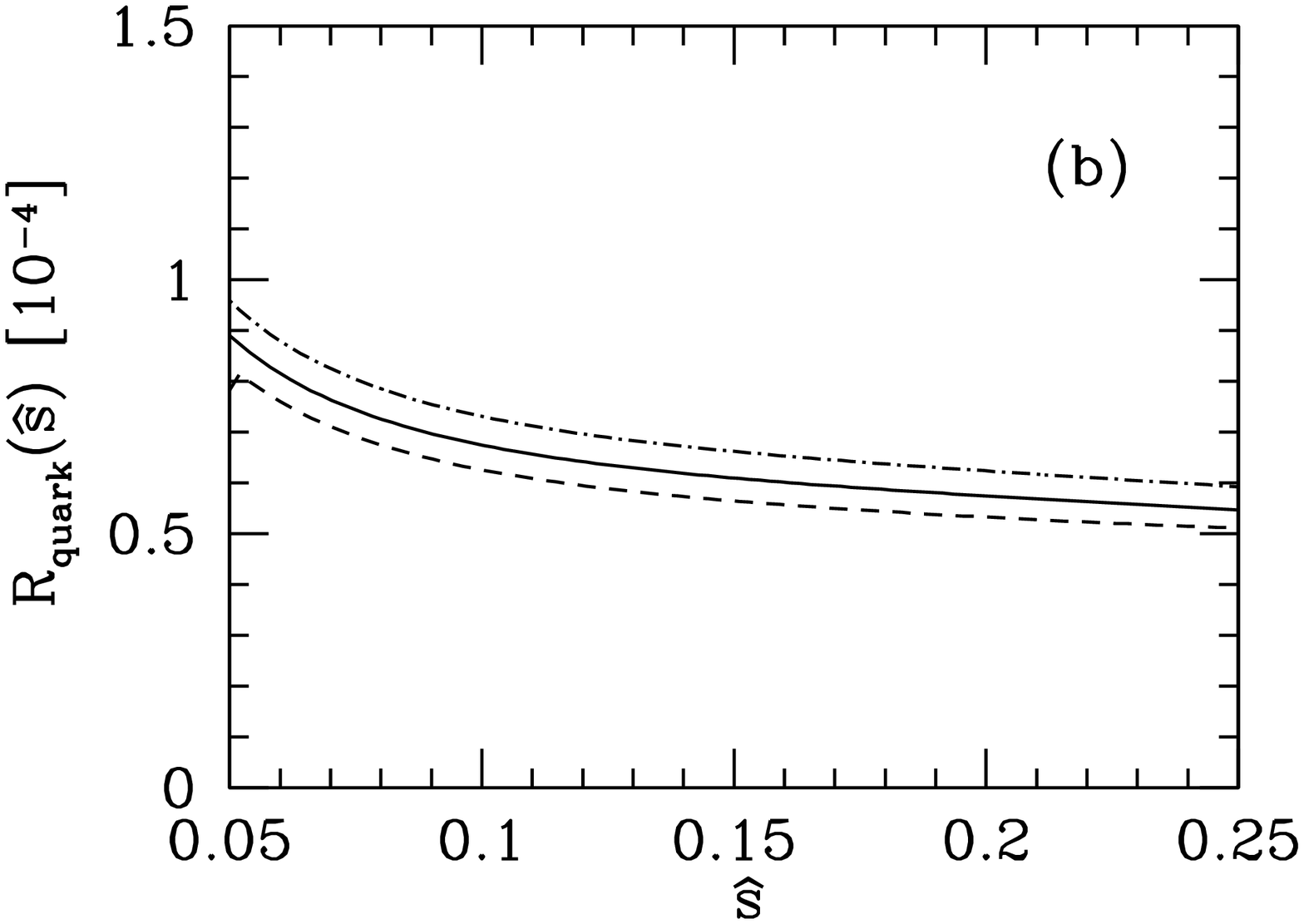}
    }
    \vspace{0.5cm}
    \caption[]{(a) The three solid lines show the $\mu$ dependence
                   of  $R_{\text{quark}}(\s)$ when
                   including the corrections to the matrix elements
                   calculated in this paper; the dashed lines 
                  are obtained when 
                   switching off
                   these corrections. We set $\hat{m}_c=0.29$.
               (b) $R_{\text{quark}}(\s)$ for 
                    $\hat{m}_c = 0.27$ (dashed line),
                    $\hat{m}_c = 0.29$ (solid line) and
                    $\hat{m}_c = 0.31$ (dash-dotted line) and $\mu$=5 GeV.
                   See text.}
\end{figure}

%
%--------------- END PLOTS ----------------
%
%--------------- BEGIN THE BIBLIOGRAPHY ---------------
%

%

\vfill
\begin{thebibliography}{99}

\bibitem{CLEOrare2}
R. Ammar et al. (CLEO Collaboration),
\Journal{\PRL} {71} {674} {1993}.

\bibitem{CLEOrare1}
R. Ammar et al. (CLEO Collaboration),
\Journal{\PRL} {74} {2885} {1995}.

\bibitem{Glenn:1998gh}
S.~Glenn et al.  (CLEO Collaboration),
%``Search for inclusive b --> s l+ l-,''
\Journal{\PRL} {80} {2289} {1998}.

\bibitem{Ligeti:1996yz}
Z.~Ligeti and M.~B.~Wise,
%``$|V_{ub}|$ from exclusive $B$ and $D$ decays,''
\Journal{\PRD} {53} {4937} {1996}.

\bibitem{Falk:1994dh}
A.~F.~Falk, M.~Luke and M.~J.~Savage,
%``Nonperturbative contributions to the inclusive rare decays
%B $\to$ X(s) gamma and B $\to$ X(s) lepton+ lepton-,''
\Journal{\PRD} {49} {3367} {1994}.

\bibitem{Ali:1997bm}
A.~Ali, G.~Hiller, L.~T.~Handoko and T.~Morozumi,
%``Power corrections in the decay rate and distributions in
%B --> X/s l+ l- in the standard model,''
\Journal{\PRD} {55} {4105} {1997} [hep-ph/9609449].

\bibitem{chen:1997} J-W.~Chen, G.~Rupak and M.~J.~Savage,
\Journal{\PLB} {410} {285} {1997}.

\bibitem{Buchalla:1998ky}
G.~Buchalla, G.~Isidori and S.~J.~Rey,
%``Corrections of order Lambda(QCD)**2/m(c)**2 
%to inclusive rare B  decays,''
\Journal{\NPB} {511} {594} {1998} [hep-ph/9705253].

\bibitem{Buchalla:1998mt}
G.~Buchalla and G.~Isidori,
%``Nonperturbative effects in anti-B --> X/s l+ l- for
%large dilepton  invariant mass,''
\Journal{\NPB} {525} {333} {1998}.

\bibitem{Krueger:1996}
F. Kr\"uger and L.M. Sehgal,
\Journal{\PLB} {380} {199} {1996}.

\bibitem{Misiak:1993bc}
M.~Misiak,
%``The b $\to$ s e+ e- and b $\to$ s gamma
% decays with next-to-leading logarithmic QCD corrections,''
 \Journal{\NPB} {393} {23} {1993}; \\
 \Journal{\NPB} {439} {461} {1995} (E).

\bibitem{Buras:1995dj}
A.~J.~Buras and M.~M\"unz,
%``Effective Hamiltonian for B $\to$ X(s)
%e+ e- beyond leading logarithms in the NDR and HV schemes,''
\Journal{\PRD} {52} {186} {1995} [hep-ph/9501281].

\bibitem{Bobeth:2000mk}
C.~Bobeth, M.~Misiak and J.~Urban,
%``Photonic penguins at two loops
% and m(t)-dependence of BR(B --> X(s) l+  l-),''
 \Journal{\NPB} {574} {291} {2000} [hep-ph/9910220].

\bibitem{Chetyrkin}
K.~Chetyrkin, M.~Misiak and M.~M\"unz,
 \Journal{\PLB} {400} {206} {1997};
 \Journal{\NPB} {518} {473} {1998};
 \Journal{\NPB} {520} {279} {1998}.

\bibitem{matching}
K.~Adel and Y.~P.~Yao,
 \Journal{\PRD} {49} {4945} {1994}; \\
C.~Greub and T.~Hurth 
 \Journal{\PRD} {56} {2934} {1997}; \\
A.~J.~Buras, A.~Kwiatkowski and N.~Pott,
 \Journal{\NPB} {517} {353} {1998}; \\
M.~Ciuchini, G.~Degrassi, P.~Gambino  and G.~F.~Giudice,
 \Journal{\NPB} {527} {21} {1998}.

\bibitem{Buchalla:1999ba}
G.~Buchalla and A.~J.~Buras,
%``The rare decays K --> pi nu anti-nu, 
%B --> X nu anti-nu and  B --> l+
\Journal{\NPB}{548}{309}{1999}
[hep-ph/9901288].

\bibitem{Grinstein:1989}
B.~Grinstein, M.~J.~Savage and M. B. Wise,
 \Journal{\NPB} {319} {271} {1989}.

\bibitem{Greub:1996tg}
C.~Greub, T.~Hurth and D.~Wyler,
%``Virtual $O(\a_s)$ corrections to the 
%inclusive decay $b \to s \gamma$,''
\Journal{\PRD} {54} {3350} {1996} [hep-ph/9603404].

\bibitem{Greub:2001sy}
C.~Greub and P.~Liniger,
%``Calculation of next-to-leading QCD corrections to b --> s g,''
\Journal{\PLB} {494} {237} {2000} [hep-ph/0008071];
\Journal{\PRD} {63} {054025} {2001} [hep-ph/0009144].

\bibitem{AAGW}
H. H. Asatryan, H. M. Asatrian, C. Greub and M. Walker,
in preparation.

\bibitem{Smirnov}
V.A. Smirnov, hep-th/9412063; \\
V.A. Smirnov, {\it Renormalization and Asymptotic Expansions},
Birkh\"auser, Basel, 1991.

\end{thebibliography}
\end{document}